\documentclass[dvips,apj]{emulateapj}
\usepackage{apjfonts,ifthen,graphicx}

\newcommand{\asca}{\textit{ASCA}}
\newcommand{\rosat}{\textit{ROSAT}}
\newcommand{\xmm}{\textit{XMM-Newton}}
\newcommand{\chandra}{\textit{Chandra}}
\newcommand{\iras}{\textit{IRAS}}
\newcommand{\fuse}{\textit{FUSE}}
\newcommand{\xspec}{\texttt{XSPEC}}
\newcommand{\wilm}{\texttt{wilm}}
\newcommand{\lodd}{\texttt{lodd}}
\newcommand{\wham}{\textit{WHAM}}

\newcommand{\nh}{$n_{\rm H}$}
\newcommand{\nhx}{$n_{{\rm H}_X}$}
\newcommand{\htwo}{H$_{2}$}
\newcommand{\etal}{et al.\@}

\submitted{submitted to the Astrophysical Journal}

\shorttitle{GALACTIC X-RAY OXYGEN ABUNDANCE}
\shortauthors{BAUMGARTNER \& MUSHOTZKY}

\begin{document}

\title{Oxygen Abundances in the Milky Way Using X-ray Absorption
Measurements\\ Towards Galaxy Clusters}

\author{W. H. Baumgartner\altaffilmark{1,2,3}}
\author{R. F. Mushotzky\altaffilmark{2}}

\altaffiltext{1}{Space Radiation Laboratory, Mail Code 220-47, 
	California Institute of Technology, 1200 E. California Bl.,
	Pasadena, CA 91125}
\altaffiltext{2}{Laboratory for High Energy Astrophysics, NASA/GSFC, 
	Code 662, Greenbelt, MD 20771}
\altaffiltext{3}{Email: {\tt wayne@srl.caltech.edu}}

\begin{abstract}

We present measurements of the oxygen abundance of the Milky~Way's ISM
by observing the K-shell X-ray photoionization edge towards galaxy
clusters.  This effect is most easily observed towards objects with
galactic columns (\nh) of a few times $10^{21}$\,cm$^{-2}$.  We
measure X-ray column densities towards 11 clusters and find that at
high galactic columns above approximately $10^{21}$\,cm$^{-2}$ the
X-ray columns are generally 1.5--3.0 times greater than the 21\,cm
\ion{H}{1} columns, indicating that molecular clouds become an
important contributor to \nh\ at higher columns.  We find the average
ISM oxygen abundance to be (O/H) $= (4.85 \pm 0.06) \times 10^{-4}$,
or 0.99 solar when using the most recent solar photospheric values.
Since X-ray observations are sensitive to the total amount of oxygen
present (gas + dust), these results indicate a high gas to dust ratio.
Also, the oxygen abundances along lines of sight through high galactic
columns (\nh) are the same as abundances through low columns,
suggesting that the composition of denser clouds is similar to that of
the more diffuse ISM.

\end{abstract}


\section{Introduction}

The measurement of the chemical abundances in our galaxy is an
important continuing area of research because knowledge of these
abundances has a significant impact on many other branches of
astronomy.  Measurements constrain theories of primordial
nucleosynthesis, and are a strong constraint on models of elemental
production in stars, amounts and composition of delayed infall, and
chemical evolution of the galaxy.

Measurements of stellar abundances are usually obtained with high
spectral resolution observations of specific optical and UV lines from
the stellar atmosphere.  Measurements of abundances in other galactic
objects such as \ion{H}{2} regions and planetary nebulae proceed along
the same lines.

The optical measurements of chemical abundances in stellar atmospheres
depend heavily on the particular model used to fit and interpret the
result, which depends on a correct determination of the ionization
balance, line blending and other physical processes like
microturbulence, granulation, and non-LTE effects.  In the case of
oxygen, a steady procession of papers
\citep{angr,grsa98,allendeprieto,wilms,asplund} has shown how the
determined solar abundance has changed substantially over time as the
models have been revised.  Also, optical measurements can typically be
made for only a few ionization states of any given element, further
complicating a determination of a total elemental abundance.

Another obstacle to determining chemical abundances in diffuse gas
(eg., the ISM, planetary nebulae, and \ion{H}{2} regions) using
optical observations is the unknown gas to dust ratio.  Since the
optical lines measured in the ISM are produced only by gas, the
fraction of the elemental abundance tied up in dust or other
ionization states is not well determined.  When this abundance for the
gas is compared to a standard solar composition, it often is less than
the sun.  This difference is usually attributed to depletion of the
element onto dust grains in the ISM, although no direct measurement of
the dust abundance was made \citep{ss96}.  Direct measurements of the
composition of the dust are difficult because they depend on
assumptions about parameters such as grain sizes and distributions
that are hard to determine.

We circumvent these problems by observing a sample of eleven galaxy
clusters with the X-ray observatory \xmm.  We use the clusters as a
background white light source against which we can see the absorption
edge at 542\,eV produced by photoionization of the inner K-shell
electrons of oxygen.  In contrast to optical observations, one can see
a signal from all the different forms of oxygen in the X-ray band.
Oxygen tied up in dust will still contribute to the absorption signal,
as well as oxygen in all of its ionization stages.  Measurements of
the K-shell oxygen edge therefore provide an excellent measure of the
total oxygen column along the line of sight, and when combined with a
measurement of the total hydrogen column can yield an oxygen abundance
determined with X-ray observations sensitive to dust and to all
ionization levels of the gas phase.

Galaxy clusters provide one of the best sets of available X-ray
sources for observing galactic abundances in absorption.  Clusters are
bright, and have relatively simple spectra at the energies of
interest.  Their spectra are dominated by continuum emission caused by
thermal bremsstrahlung in the hot intracluster plasma.  They are
extragalactic, occur at all galactic latitudes, and their emission
does not change appreciably on terrestrial time scales.  They are
optically thin and have no intrinsic absorption to complicate galactic
measurements.  Further, a few bright clusters are objects suitable for
observation by high resolution gratings; however, we limit ourselves
to X-ray CCD imaging spectroscopy in this paper in order to obtain the
largest possible uniform sample.

The \xmm\ satellite is the ideal instrument for this purpose because of
its very high sensitivity and good CCD spectral resolution.  Also, its
low energy broad band response below 2\,keV has a fairly well
understood calibration, which has not always been the case for X-ray
telescopes at this energy.

\subsection{ISM Observations}

Until recently, observations of the Milky Way ISM often showed
subsolar abundances.  Observations by \cite{fitz} using absorption
lines towards halo stars with the GHRS have shown that the galactic
elemental abundances were subsolar.  Subsolar abundances have also
been reported for oxygen explicitly \citep{meyer,cardelli96}.  These
observations, along with others (such as the one by \cite{cardelli94}
that found the ISM krypton abundance to be 60\% of the solar value),
led to the idea that the ISM abundances were approximately \twothirds\
the solar values \citep{mathis}.

\cite{sofia} later summarized these developments and showed that the
ISM oxygen data were more consistent with the new lower solar oxygen
abundances \citep{holweger,asplund}.  Recent data from \fuse\
observations of many ISM sightlines \citep{andre,jensen} show that the
ISM gas phase oxygen abundance is (O/H)$_{\mbox{\scriptsize{gas}}} =
4.08 \times 10^{-4}$ and $4.39 \times 10^{-4}$, respectively.  This is
close to the solar value from \cite{asplund} of $4.79 \times 10^{-4}$,
and supports only mild depletion of oxygen in the ISM.  Other \fuse\
results from \cite{oliveira} support a lower gas phase oxygen
abundance of $3.63 \times 10^{-4}$ along the lines of sight towards
four white dwarfs.

X-ray observations of absorption in the ISM have produced similar
results.  \cite{ab} used the PSPC on \rosat\ to measure the X-ray
hydrogen column \nhx\ towards 26 clusters and found that the X-ray
column exceeds the 21\,cm column for columns above $5 \times
10^{20}$\,cm$^{-2}$ and that extra absorption from molecular hydrogen
is required.  Higher resolution grating observations of the ISM with
\chandra\ \citep{devries,juett} have started to reveal the structure
of the oxygen K-edge and put constraints on the oxygen ionization
fraction.  \cite{juett} have found that the ratio of
\ion{O}{2}/\ion{O}{1} $\approx$ 0.1, and that the precise energy of
the gas phase oxygen K-edge is 542\,eV.

The resolution of the grating observations from \chandra\ and \xmm\
are unsurpassed, and allow for a very careful determination of the
absorbing galactic oxygen column towards background sources
\citep{paerels,juett,devries,page}.  However, the hydrogen column
towards these sources is often not measured directly and as a result
the oxygen abundance is not obtained.  \cite{paerels} using the
\chandra\ LETGS did derive an equivalent \nh\ from the overall shape
of the spectrum towards the galactic X-ray binary X0614+091, and
obtained an oxygen abundance of 0.93 solar on the \cite{wilms} solar
abundance scale.  \cite{weisskopf} performed a similar measurement
towards the Crab with the \chandra\ LETGS and obtained a galactic
oxygen abundance of 0.68 solar on the Wilms scale.  \cite{willingale}
has used the MOS CCD detectors onboard \xmm\ to measure the abundance
towards the Crab and finds an oxygen abundance of 1.03 solar (Wilms), and
\cite{vuong} presents measurements towards star forming regions that
are also best fit with the new solar abundances.

\subsection{Solar Abundances}

There has been some controversy in the literature as to the canonical
values to use for the solar elemental abundances.  X-ray absorption
observations give directly the column density of each element having
its K-edge in the bandpass.  When combined with a hydrogen column
density, this yields an elemental abundance by number with respect to
hydrogen.  However, for the sake of convenience elemental abundances
are often reported with respect to the solar values.

The compilation of \cite{angr} has been a standard for this purpose.
They published abundances for the natural elements compiled from
observations of the solar photosphere and from measurements of
primitive CI carbonaceous chondritic meteorites.  For many of the
elements presented, there was a good agreement between the meteoritic
and photospheric values for elements where both types can be measured.
However, there was still a discrepancy for some important elements
such as iron.

Since 1989, the situation has improved.  Reanalysis of the stellar
photospheric data for iron that includes lines from \ion{Fe}{2} in
addition to \ion{Fe}{1} as well as improved modeling of the solar
lines \citep{grsa99} have brought the meteoritic and photospheric
values into agreement. \citet{grsa98} incorporate these changes and
others.  Of more importance to our work on the galactic oxygen
abundance, the solar abundances of carbon, nitrogen, and oxygen have
also changed since the compilation of \cite{angr}.  Measurements of
these elements in the sun cannot be easily reconciled with meteoritic
measurements because they form gaseous compounds easily and are found
at much lower abundances in the CI meteorites than in the sun.
\cite{holweger,allendeprieto,asplund} have made improvements to the
solar oxygen abundance by including non-LTE effects, using three
dimensional models, deblending unresolved lines, and incorporating a
better understanding of solar granulation on the derived measurements.
The carbon \citep{allendeprieto} and nitrogen \citep{holweger}
abundances have also improved in a similar fashion, as reported by
\cite{lodders}.  The \cite{wilms} compilation available in the X-ray
software package \xspec\ has solar abundance values for carbon,
nitrogen, and oxygen (O/H$= 4.90 \times 10^{-4}$ by number) consistent
with the most recent values.  These downward revisions in the solar
photospheric oxygen abundance have allowed ISM oxygen measurements to
finally agree with the solar standard.

\section{X-ray Observations}

\subsection{Sample Selection}

Our sample of 11 galaxy clusters was chosen from the public archives
of the \xmm\ satellite.  The main criteria for selection are that the
cluster has a 21\,cm galactic hydrogen column greater than $0.5 \times
10^{21}\,\mbox{cm}^{-2}$ and that the observation have more than $5
\times 10^4$ counts in the EPIC spectrum in order to ensure a good
measurement of the absorption from galactic oxygen.

\begin{figure}[!t]
\resizebox{\columnwidth}{!}{\rotatebox{+90}{\includegraphics{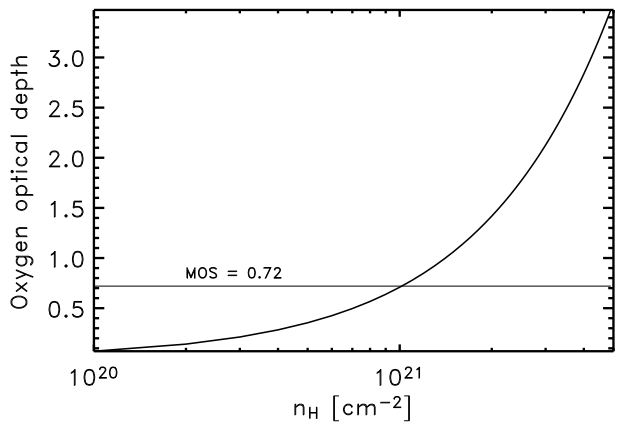}}}
\caption[The Mily Way ISM optical depth in oxygen K-shell absorption as a
  function of the hydrogen column density for an ISM with standard
  solar abundances as given in \cite{wilms}]{The MW ISM optical depth
  in oxygen K-shell absorption as a function of the hydrogen column
  density for an ISM with standard solar abundances as given in
  \cite{wilms}.  The optical depth of the oxygen edge in the MOS
  detectors resulting from filters, windows, etc.\ is also plotted for
  comparison and is $\tau = 0.72$.
\label{oxy_tau_nh}}
\end{figure}

The choice of \nh\ greater than $0.5 \times 10^{21}\,\mbox{cm}^{-2}$
was set by the need to have galactic oxygen optical depths near $\tau
= 1$ in order to allow good measurements of the oxygen abundance.
Figure~\ref{oxy_tau_nh}
shows how the galactic oxygen optical depth is related to \nh\ given
an ISM with the solar abundances of \cite{wilms}.  An oxygen optical
depth of 1.0 is reached at a hydrogen column of approximately $1.5
\times 10^{21}\,\mbox{cm}^{-2}$.

\begin{deluxetable*}{lrrrrrrrrrrrrr}[!tb]
\tablecaption{Cluster Reference Values\label{table_reference}}
\tablewidth{0pt}
\tablehead{
\colhead{Cluster} & \colhead{RA} & \colhead{dec} & \colhead{l} & \colhead{b} & \colhead{$n_H$\tablenotemark{a}} & \colhead{$A_B$\tablenotemark{b}} & \colhead{$E(B-V)$\tablenotemark{c}} &\colhead{\textit{IRAS}\tablenotemark{d}} &\colhead{$kT$} & \colhead{$Z$\tablenotemark{e}} &\colhead{$z$} & \colhead{Exposure} & \colhead{\textit{XMM}} \\& \colhead{[J2000.0]} & & & & & & & 100\,$\mu$m & \colhead{[keV]} & & & \colhead{[ksec]} & \colhead{Rev.}
}
\startdata
   PKS\,0745-19 & 116.883 & -19.296 & 236.444 &   3.030 & 4.24 &  2.252 &  0.522 &  36.9 &   6.25 & 0.61 &  0.103 & 17.61 & 164\\
     ABELL\,401 &  44.737 &  13.582 & 164.180 & -38.869 & 1.05 &  0.678 &  0.157 &  10.2 &   8.07 & 0.49 &  0.074 & 12.80 & 395\\
       Tri\,aus & 249.585 & -64.516 & 324.478 & -11.627 & 1.30 &  0.592 &  0.137 &   7.7 &  10.19 & 0.45 &  0.051 & 9.40 & 219\\
           AWM7 &  43.634 &  41.586 & 146.347 & -15.621 & 0.98 &  0.504 &  0.117 &   5.5 &   3.71 & 0.89 &  0.018 & 31.86 & 577\\
     ABELL\,478 &  63.336 &  10.476 & 182.411 & -28.296 & 1.51 &  2.291 &  0.531 &  18.1 &   7.07 & 0.54 &  0.088 & 46.77 & 401\\
RX\,J0658.4-5557 & 104.622 & -55.953 & 266.030 & -21.253 & 0.65 &  0.335 &  0.078 &   4.6 &  11.62 & 0.28 &  0.296 & 21.05 & 159\\
    ABELL\,2163 & 243.892 &  -6.124 &   6.752 &  30.521 & 1.21 &  1.528 &  0.354 &  21.8 &  12.12 & 0.38 &  0.203 & 10.99 & 132\\
     ABELL\,262 &  28.210 &  36.146 & 136.585 & -25.092 & 0.54 &  0.373 &  0.086 &   4.3 &   2.17 & 0.87 &  0.016 & 23.62 & 203\\
   2A\,0335+096 &  54.647 &   9.965 & 176.251 & -35.077 & 1.78 &  1.771 &  0.410 &  18.9 &   2.86 & 1.01 &  0.035 & 1.81 & 215\\
     ABELL\,496 &  68.405 & -13.246 & 209.568 & -36.484 & 0.46 &  0.586 &  0.136 &   4.0 &   3.89 & 0.82 &  0.033 & 30.14 & 211\\
     CIZA\,1324 & 201.180 & -57.614 & 307.394 &   4.969 & 3.81 &  3.164 &  0.733 &  36.8 &   3.00 & 0.87 &  0.019 & 10.76 & 675\\

\enddata
\tablenotetext{a}{Values are the hydrogen column [$10^{21}$ cm$^{-2}$]
  from the 21\,cm work of \cite{dl}. \cite{ab} estimate the errors to
  be 5\% on these data.}
\tablenotetext{b}{From \cite{schlegel}.}
\tablenotetext{c}{From NED.}
\tablenotetext{d}{Values in MJy sr$^{-1}$ taken from the all sky maps of \cite{schlegel}.}
\tablenotetext{e}{Total metal abundance from \cite{horner01}, rescaled from the \cite{angr} to the \cite{wilms} solar abundances.}
\end{deluxetable*}

Reference data for these clusters can be found in
Table~\ref{table_reference}.  We include the cluster coordinates, the
\nh\ column from the 21\,cm work of \cite{dl}, the optical extinction,
color excess, \textit{IRAS} 100\,$\mu$m count, previously determined
X-ray values for the cluster temperature and metallicity from the
\asca\ observations in \citet{horner01} and \citet{horner03}, the
optical redshift, the length of the \xmm\ exposure, and the \xmm\
orbit number.

\begin{figure*}[!t]
\resizebox{\textwidth}{!}{\rotatebox{90}{\includegraphics{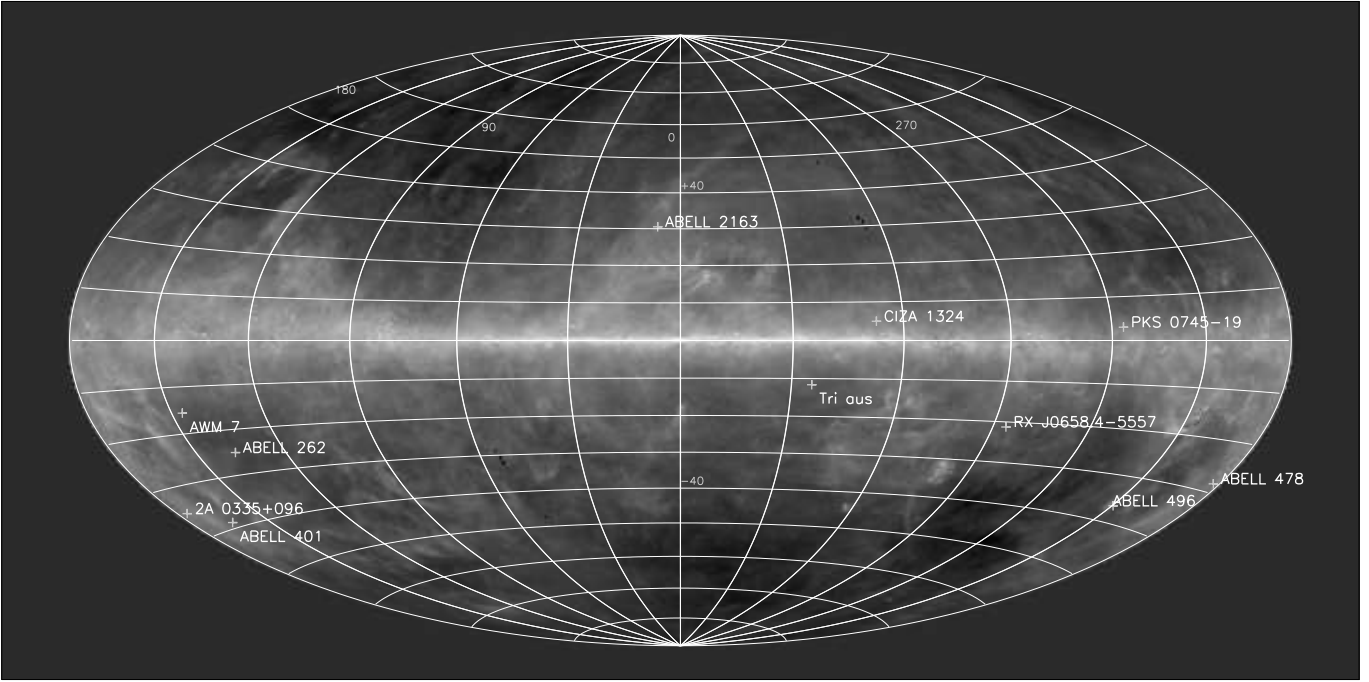}}}
\caption[A map showing the location of the clusters in the ISM
  absorption sample]{A map showing the location of the clusters in our
  sample.  Each cluster is shown plotted on a full sky map in galactic
  coordinates, with the galactic center at the figure center.  The
  greyscale image is the \iras\ 100\,$\mu$m map \citep{schlegel}, a
  good tracer for dust in the galaxy.
\label{cluster_map}}
\end{figure*}

The requirement of a high \nh\ places most of our sample near the
galactic plane.  Figure~\ref{cluster_map}
shows the location of each of the clusters in our sample superimposed
upon the all sky map of the 100\,$\mu$m emission observed by \iras.
The \iras\ map is a good tracer of dust in the galaxy, and shows
that the clusters in our sample that do not lie in the galactic plane
are still located in areas with high \nh.  Figure~\ref{cluster_map}
also shows that our clusters sample a wide range of galactic sight lines.

\subsection{Data Reduction}

We extracted spectra from the pn and two MOS CCD detectors that
comprise the EPIC camera on \xmm.  These detectors' moderate
resolution of $\Delta E \sim 50--100$\,eV and large field of view (30
arcminute diameter) are well matched to our requirements.  The cluster
data were re-extracted from the raw ODF files using SAS version 5.4.1.
After filtering to eliminate periods with high background rates, we
choose regions on the CCDs that encompass most of a cluster's emission
without a significant contribution from the background.  This was done
by selecting by eye a circular region centered on the cluster with a
radius extending to where the cluster emission drops to about three
times the background level.  Backgrounds were local, and taken from
areas in the field of view without cluster emission.  Ancillary
response files and response matrices are generated using {\tt{arfgen}}
and {\tt{rmfgen}} within SAS.

We extracted spectra between 0.46--10.0\,keV in the MOS1 and MOS2
detectors, and between 0.46--7.2\,keV in the pn in order to avoid
large background lines above 7.2\,keV in the pn.  Our lower energy
cutoff is 0.46\,keV in order to avoid problems with the calibration of
the redistribution function at very low energies\footnote{see
http://xmm.esac.esa.int/docs/documents/CAL-SRN-0169-1-0.ps.gz}.

\subsection{Extra Edge}
\label{extraedge}

\begin{figure}[!t]
\resizebox{\columnwidth}{!}{\rotatebox{-90}{\includegraphics{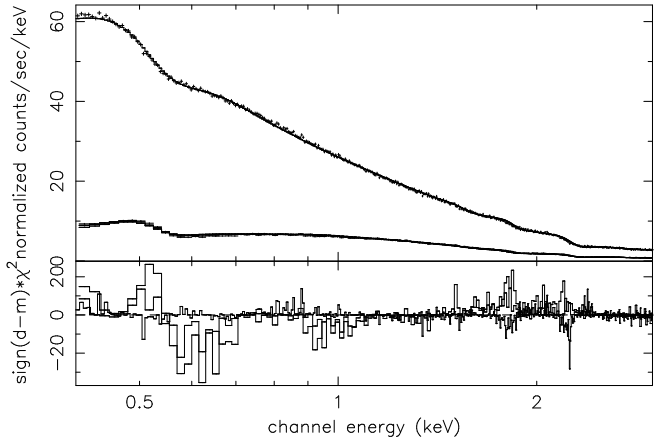}}}
\caption[A spectral fit to the bright quasar 3C~273 with the 3 CCD
detectors of the EPIC camera on \xmm]{A spectral fit to the bright
quasar 3C~273 with the 3 CCD detectors of the EPIC camera on \xmm.
The residuals above the oxygen edge at 544\,eV show the large
discrepancy between detectors; the higher resolution data from the pn
detector is well fit throughout the band, while the MOS data have
significant residuals.  We address this problem by including an extra
edge to the model that is discussed in section \S\ref{extraedge}.
\label{3c273fit}}
\end{figure}

In order to obtain the greatest signal to background ratio we would
like to fit the data from all three CCD detectors on \xmm.  However,
early on we discovered that oxygen absorption results from the three
detectors did not agree.  For some clusters, when the MOS detectors
showed substantial absorption the pn detector showed almost none.  We
examined data from an \xmm\ observation of the bright quasar 3C~273 (a
good continuum source at the energies of interest) in order to
investigate this effect at low galactic columns (3C~273 has \nh\ =
$1.8 \times 10^{20}\,\mbox{cm}^{-2}$).  We found that there are
substantial residuals in the MOS detectors at the location of the
oxygen edge.  Figure~\ref{3c273fit}
shows the residuals to the 3C~273 fit at the location of the oxygen
edge and illustrates the large deviation in the MOS detectors.  We
also observe this effect towards the Coma cluster, another low
galactic column source (\nh\ = $9.2 \times 10^{19}\,\mbox{cm}^{-2}$).

\begin{figure}[!t]
\resizebox{\columnwidth}{!}{\rotatebox{-90}{\includegraphics{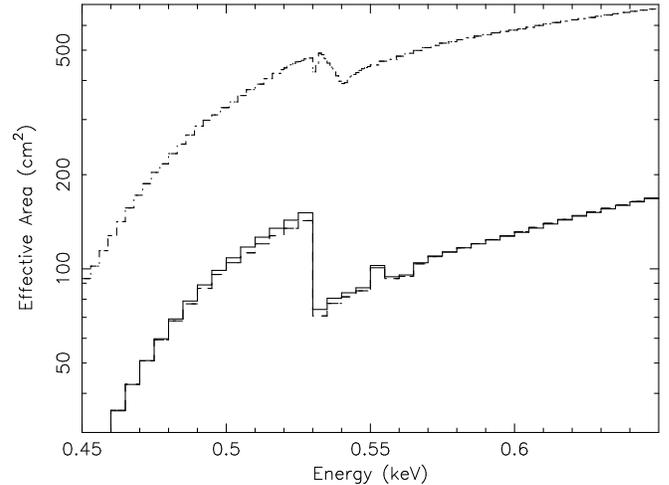}}}
\caption[The instrumental response of the MOS detector in the region
  of the oxygen edge]{The instrumental response of the detectors of
  the EPIC camera in the region of the oxygen edge.  The upper trace
  is the pn detector, and the lower two traces are the MOS detectors.
  The optical depth of the MOS instrumental edge is $\tau = 0.72$.
  The different shape in the edge structure of the front-side
  illuminated MOS and back-side illuminated pn is clearly seen.
\label{mos_edge}}
\end{figure}

We interpret this effect as an inaccuracy in the response matrices for
the MOS detectors.  Such an effect could possibly be caused by the
outgassing of organic materials onto the surface of the MOS detectors,
and is seen in observations taken with both the thin and medium
filters.  The oxygen edge in the detector is caused by molecular
compounds of oxygen, and its energy of 0.53\,keV is slightly offset
from the ISM atomic edge at 0.542\,keV\@.  However, the correct
determination of this instrumental feature could effect measurements
of the ISM oxygen abundance at low column densities.
Figure~\ref{mos_edge} shows the magnitude of the instrument edge in
the MOS detectors.

We compensate for this problem by introducing an extra edge into the
fit at the solid state oxygen K-shell energy of 0.53\,keV\@.  When the
extra edge component has optical depths of 0.22 for the MOS1 and 0.20
for the MOS2 the fitted oxygen absorption is consistent between all
three detectors for 3C~273 and Coma, so we use these values for our
cluster fits. 

This problem has been brought to the attention of the \xmm\ EPIC
instrument team.  They have kindly supplied a beta version of a
revised quantum efficiency file that is meant to address this
problem.  The tests we have made using this file have provided results
that are consistent with the extra edge method described above.

\subsection{Spectral Fitting}

We use \xspec\ version 11.3.0 to fit the data.  The cluster emission
is modeled by the {\tt{apec}} (v.\@ 1.3.1) plasma code \citep{apec},
and the intervening galactic material by the {\tt{tbvarabs}} model
\citep{wilms} (\xspec\ model: {\tt{edge * tbvarabs * apec}}; the
edge component of the model is discussed in \S\ref{extraedge}).

The {\tt{tbvarabs}} model models the absorption due to photoionization
from the abundant elements up to nickel.  It also takes into account
absorption from molecular hydrogen and depletion of the elements onto
grains.  The abundance of each of the elements can be individually
fit, as well as the depletion fraction.  The model elemental
abundances are computed from the observables (e.g., spectral line
equivalent width, optical depth) by specifying a table of solar
elemental abundances selectable by the user.  We choose the
compilation of theoretical X-ray absorption cross sections by
\cite{verner}\footnote{implemented with the \xspec\ command:
{\tt{xsect vern}}}, and the solar abundances of
\cite{wilms}\footnote{implemented with the \xspec\ command: {\tt{abund
wilm}}}.  Although \cite{lodders} has published a more recent
compilation, we feel that her solar abundance for helium that takes
into account heavy element settling in the sun results in an abundance
that is too low for good ISM modeling.  Her compilation of recent
results for carbon, nitrogen, and oxygen needs to be taken into
account and is very similar to the Wilms~\etal\ model abundances that
we use.

For our investigation, the dominant contributors to galactic
absorption are hydrogen, helium, and oxygen.  Secondary contributors
observable in the \xmm\ band are neon and iron (from the L-shell).  We
initially set the hydrogen column to the galactic 21\,cm value from
\cite{dl}, but allow it to vary with the fit.  The helium abundance is
not well constrained independently of the hydrogen value, and is set
to solar in our fit.  Neon and iron are also initially set to their
solar abundance values, but are allowed to vary with the fit.  All
other elements are fixed to their solar abundance values as determined
by \cite{wilms}.  The grain and depletion parameters are set to their
default value in the {\tt{tbvarabs}} model, but these parameters do
not significantly affect the results of our fitting.  All of the
parameters in the {\tt{tbvarabs}} portion of the model are constrained
to have the same value for each of the three EPIC detectors.

Initial values for the cluster temperature, metal abundance, and
redshift are given in Table~\ref{table_reference}, taken from the work
of \cite{horner01}.  The cluster temperature and metal abundance are
constrained to have the same value for the three EPIC detectors, but
the cluster redshift is allowed to vary separately in each detector in
order to compensate for any small energy calibration errors in the
data.

The clusters Abell~478 and Abell~262 had high $\chi^2$ values compared
to the other clusters when fit with one temperature plasma models.
This is a result of a lower temperature cooling flow component in the
cluster contributing flux at lower energies and causing the model to
underestimate the galactic hydrogen and oxygen absorption.
\cite{kaastra} show that Abell~262 is more successfully fit with a two
temperature model using the same \xmm\ observations that we have
extracted from the archives for this paper, and \cite{sakelliou} also
require a two temperature model for Abell~478.  For these clusters, we
use a two component thermal model with galactic absorption and an
extra edge, {\tt [edge*tbvarabs*(apec + apec)]} and quote the
temperature of the higher temperature component representing the main
cluster emission in Table~\ref{table_results}.  The metal abundance of
the second thermal component was tied to that of the first component
and the normalization left free.

\begin{figure}
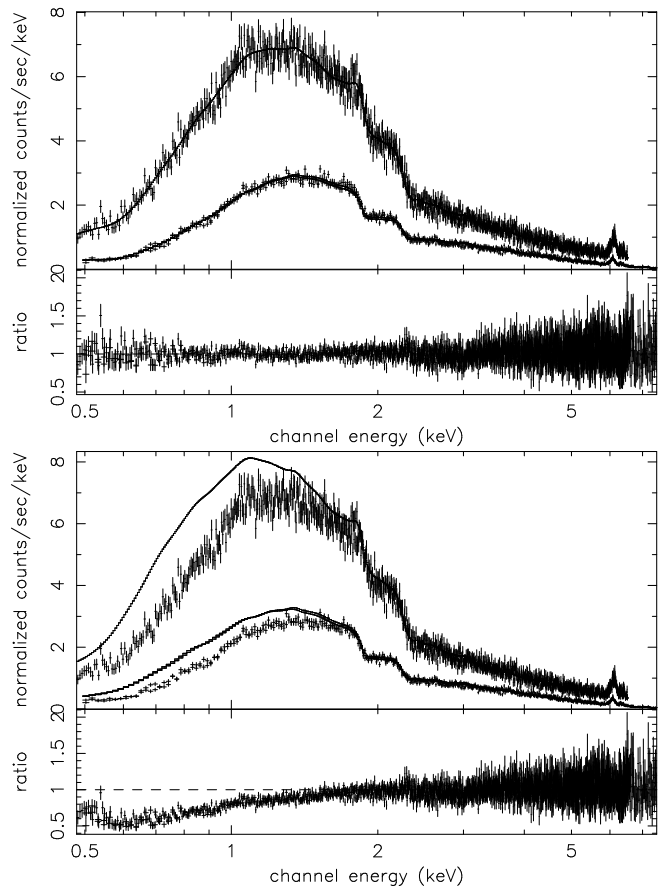

\begin{center}
\resizebox{\columnwidth}{!}{\rotatebox{-90}{\includegraphics{fig5.ps}}}
\resizebox{\columnwidth}{!}{\rotatebox{-90}{\includegraphics{fig6.ps}}}
\end{center}
\caption{Fits to the cluster PKS~0745-19.  The data are shown with
  errors and the model is a solid line in the top plot of each panel.
  The residuals are shown in the bottom plot of each panel.  The top
  panel shows that the overall fit is good to a model that includes an
  extra edge component in the MOS detectors, as is the .5\,keV region
  where the oxygen signal lies.  The biggest residual is at 1\,keV and
  is from iron L-shell emission, and does not affect the oxygen fit.
  In the bottom panel, the data was first fit with the standard
  model. Then, the model oxygen abundance was set to zero to
  illustrate the strength and importance of the galactic oxygen
  absorption component.
\label{cluster_fit}}
\end{figure}

A spectrum with residuals to the fit for a typical cluster is given in
Figure~\ref{cluster_fit}.
In the top panel we show the spectrum fit with the model, while in the
bottom panel we set the galactic oxygen abundance to zero to
illustrate the strength of the absorption signal.
 
\section{Results}

\renewcommand{\arraystretch}{1.5}
\begin{deluxetable*}{lrrrrrrrr}[!tb]
\tablecaption{X-ray Determined Galactic Absorption towards Galaxy Clusters\label{table_results}}
\tablewidth{0pt}
\tablehead{
\colhead{Cluster} & \colhead{$z$} & \colhead{$kT$} & \colhead{Z\tablenotemark{a}} & \colhead{\nh\tablenotemark{b}} & \colhead{Oxygen\tablenotemark{c}} & \colhead{Counts\tablenotemark{d}} & \colhead{$\chi^2$} &\colhead{dof} \\& & \colhead{[keV]} & & & \colhead{abundance} & & \colhead{per dof} &
}
\startdata
    PKS 0745-19 & 0.100 &  7.173 & 0.68 & 5.620$_{5.350}^{5.930}$ & 0.74$_{0.59}^{0.88}$ & 3.7e+05 &  1.09 & 2121\\
     ABELL 0401 & 0.069 &  8.738 & 0.58 & 1.050$_{0.990}^{1.080}$ & 0.56$_{0.40}^{0.79}$ & 2.8e+05 &  1.06 & 1918\\
        Tri aus & 0.050 & 10.144 & 0.57 & 1.500$_{1.430}^{1.570}$ & 1.02$_{0.86}^{1.21}$ & 4.2e+05 &  1.05 & 2202\\
           AWM7 & 0.016 &  3.652 & 1.14 & 1.140$_{1.130}^{1.160}$ & 0.95$_{0.92}^{1.02}$ & 1.4e+06 &  1.25 & 2362\\
     ABELL 0478\tablenotemark{e} & 0.081 &  6.588 & 0.66 & 3.640$_{3.630}^{3.660}$ & 1.03$_{1.01}^{1.06}$ & 1.9e+06 &  1.39 & 2348\\
RX J0658.4-5557 & 0.297 & 12.392 & 0.44 & 0.310$_{0.250}^{0.370}$ & 0.48$_{0.00}^{1.15}$ & 1.5e+05 &  0.99 & 1474\\
     ABELL 2163 & 0.194 & 12.997 & 0.40 & 2.130$_{2.000}^{2.290}$ & 0.97$_{0.74}^{1.21}$ & 1.1e+05 &  1.02 & 1428\\
     ABELL 0262\tablenotemark{e} & 0.014 &  2.057 & 0.78 & 0.850$_{0.790}^{0.900}$ & 0.70$_{0.43}^{0.85}$ & 4.4e+05 &  1.26 & 1727\\
    2A 0335+096 & 0.034 &  2.661 & 0.98 & 3.160$_{2.960}^{3.350}$ & 0.82$_{0.69}^{1.10}$ & 5.8e+04 &  1.18 &  936\\
     ABELL 0496 & 0.031 &  3.556 & 0.96 & 0.640$_{0.620}^{0.660}$ & 0.76$_{0.68}^{0.86}$ & 5.8e+05 &  1.29 & 2059\\
      CIZA 1324 & 0.018 &  2.944 & 0.94 & 5.750$_{5.180}^{6.500}$ & 1.32$_{0.95}^{1.73}$ & 8.8e+04 &  1.09 & 1220\\

\enddata
\tablecomments{All values are from the X-ray fit to the \xmm\ data.  Values in sub and superscript are the range of the 90\% confidence interval.}
\tablenotetext{a}{Cluster metal abundance with respect to \cite{wilms}.}
\tablenotetext{b}{Total hydrogen column in units of 10$^{21}$\,cm$^{-2}$.}
\tablenotetext{c}{The oxygen abundance of the galactic absorption component is given with respect to the solar value of \cite{wilms}.}
\tablenotetext{d}{Total counts in the EPIC detectors after background particle filtering.}
\tablenotetext{e}{The model fit to the data for this cluster includes
     an extra thermal component representing emission froma a cooling
     flow as noted in the text.}
\end{deluxetable*}
\renewcommand{\arraystretch}{1.0}

The results from the X-ray spectral fits are shown in
Table~\ref{table_results}.  

Column~1 gives the cluster name, while columns 2--4 give the fitted
cluster redshift, temperature, overall metal abundance, and hydrogen
column.  We give the results for the observed X-ray column of galactic
hydrogen in column~5 of the table.  The main results are for the
elemental abundance of galactic oxygen along the line of site towards
the background cluster and are given in column~6.  Columns~7, 8, and 9
of the table are meant to illustrate the quality of the fit and list
the number of photons in the X-ray spectrum from all three EPIC
detectors, the reduced $\chi^2$, and the number of degrees of freedom
of the fit.

\subsection{Hydrogen}

The total galactic hydrogen column density is composed of several
parts: the neutral atomic gas measured by 21\,cm radio observations;
the warm, ionized \ion{H}{2} gas that is sometimes associated with
H$\alpha$ emission; and molecular hydrogen, \htwo, often associated
with CO emission.  The X-ray measure of the hydrogen column, \nhx, is
sensitive to all these forms and is indicative of the total hydrogen
column.

The X-ray total column is also sensitive to the helium component along
the line of sight.  At the X-ray energies of these observations, the
absorption cross section due to helium is substantial and is greater
than the hydrogen component.  However, the helium is mostly primordial
in origin and the hydrogen to helium ratio in the galaxy is assumed to
be uniform.  Therefore, variations in the helium abundance or
distribution are not expected to have a significant effect on the
total hydrogen columns.  

\begin{figure}[!t]
\resizebox{\columnwidth}{!}{\rotatebox{+90}{\includegraphics{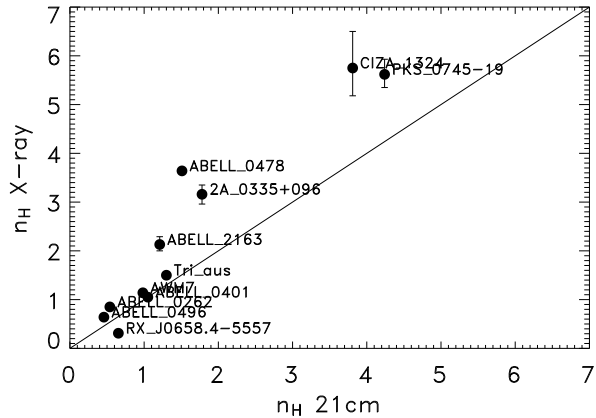}}}
\caption[The X-ray total hydrogen column density plotted against the
neutral hydrogen value derived from the 21\, cm observations in
\cite{dl}]{The X-ray total hydrogen column density plotted against the
neutral hydrogen value derived from the 21\, cm observations in
\cite{dl}.  The hydrogen columns are in units of $10^{21}$
cm$^{-2}$. Many clusters have error bars smaller than the plotted
points.  The X-ray values are generally higher because these
observations are sensitive to all forms of hydrogen and detect \htwo\
that could not be seen with the 21\,cm observations.  The molecular
component of \nh\ becomes most important above columns of $\sim 0.5
\times 10^{21}$\,cm$^{-2}$ where the \htwo\ is sufficiently dense
enough to shield itself from dissociating radiation from hot young
galactic stars.
\label{nh_nh}}
\end{figure}

\begin{figure*}[!tb]
\resizebox{\textwidth}{!}{\rotatebox{+90}{\includegraphics{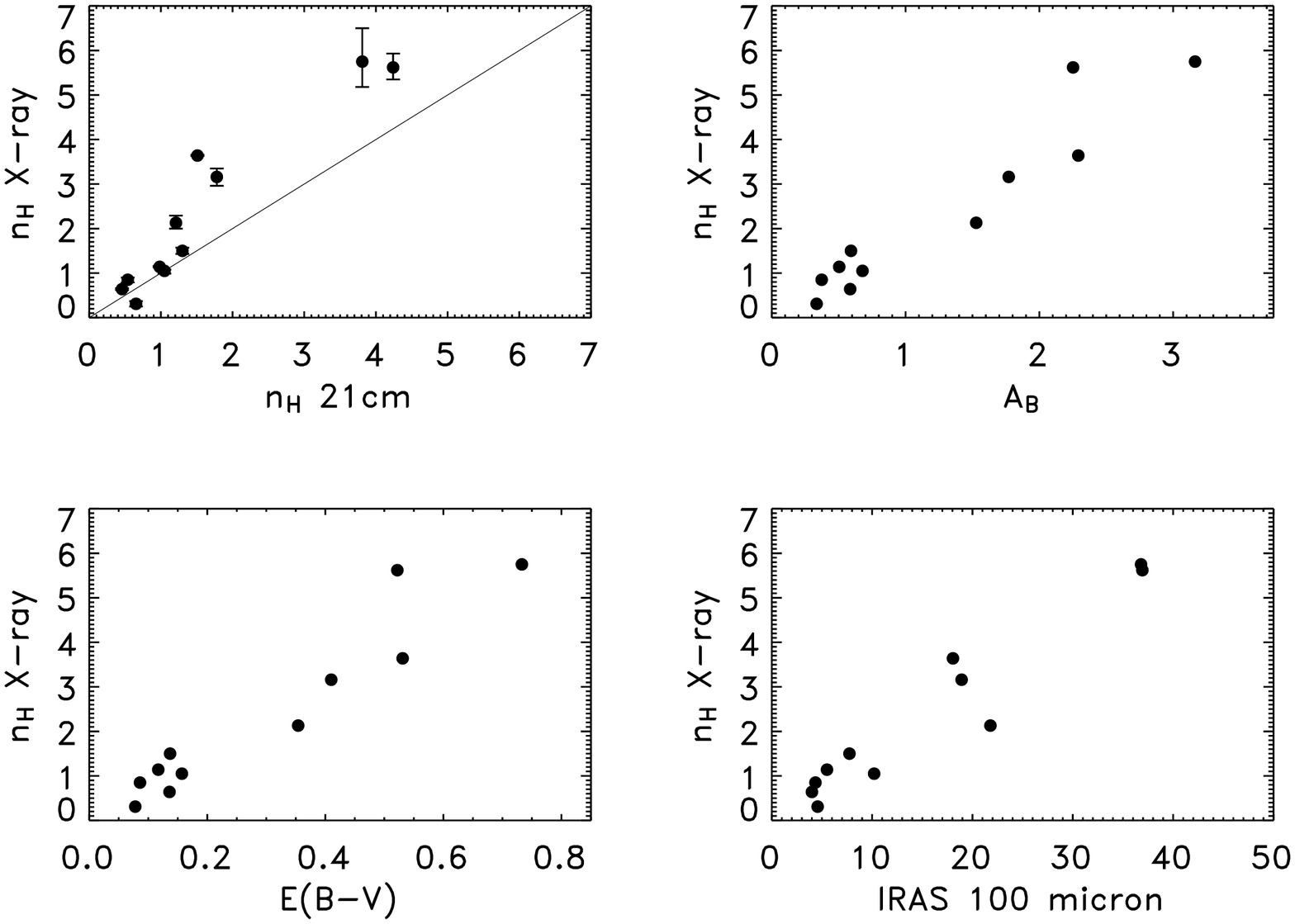}}}
\caption[The correlation of \nh\ derived from X-ray observations with
  other measures of the ISM towards the lines of sight of the 11
  clusters in our sample]{The correlation of \nh\ derived from X-ray
  observations with other measures of the ISM towards the lines of
  sight of the 11 clusters in our sample.  The top left panel shows
  \nhx\ against the 21\,cm radio observation of neutral atomic
  hydrogen [the data are in units of $10^{21}$\,cm$^{-2}$]; the top
  right panel shows shows \nhx\ against the $B$-band optical
  extinction data in magnitudes; the lower left panel shoes \nhx\
  against the optical reddening; and the lower right panel shoes \nhx\
  against the \iras\ DIRBE corrected $100\mu$m emission in
  MJy\,sr$^{-1}$.  The X-ray emission is better correlated with the
  reddening and \iras\ emission which are also sensitive to the
  effects of dust and molecular clouds.
\label{compare}}
\end{figure*}

\begin{figure}[!t]
\resizebox{\columnwidth}{!}{\rotatebox{+90}{\includegraphics{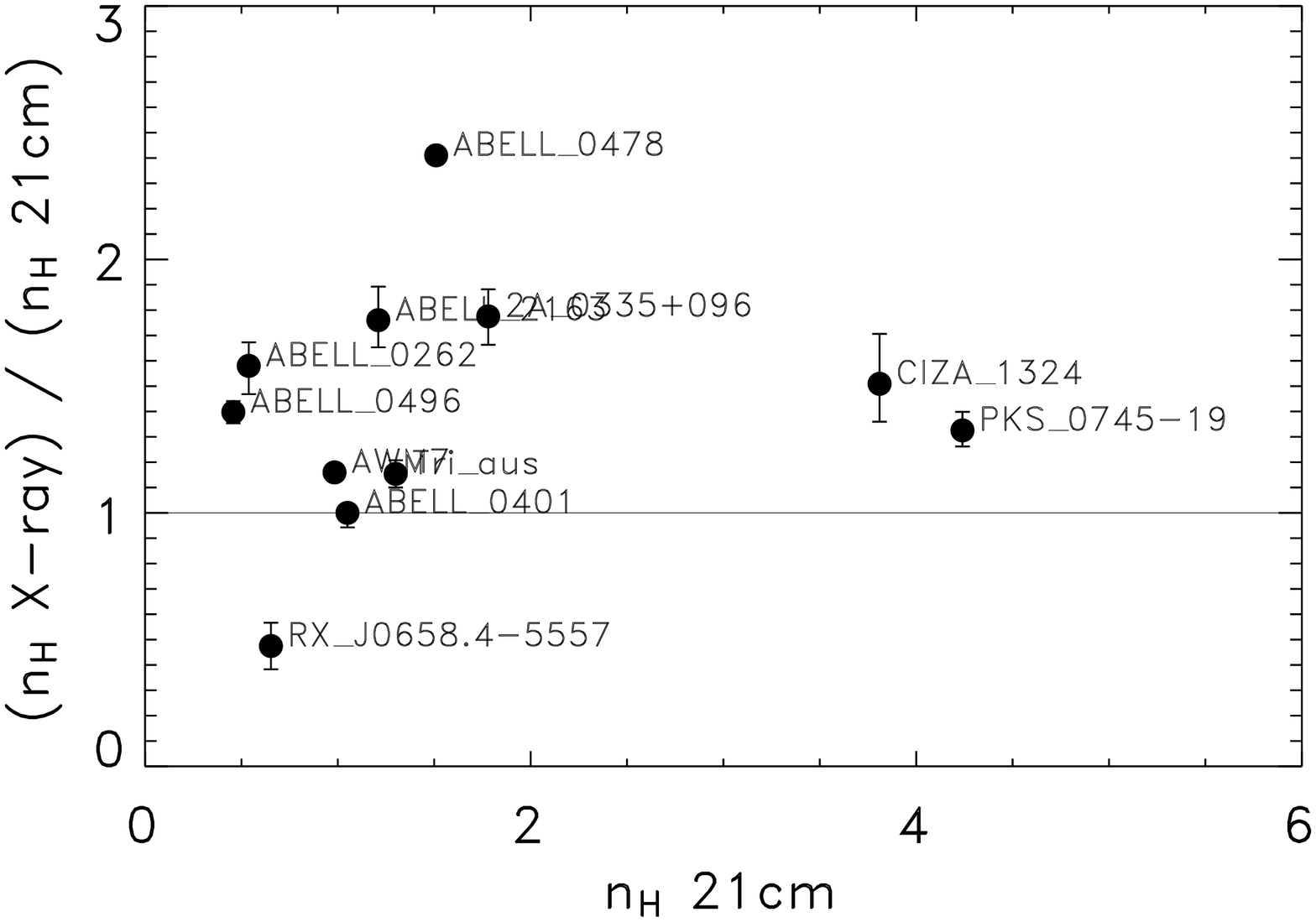}}}
\caption{The overabundance of the X-ray derived \nh\ compared to the
  21\,cm derived value [in units of $10^{21}$\,cm$^{-2}$].
\label{nh_nh_ratio}}
\end{figure}

Figure~\ref{nh_nh} 
shows the X-ray determined total hydrogen column from this work
plotted against the column of neutral hydrogen measured by the 21\,cm
observations of \cite{dl}.  The X-ray measure of the hydrogen column
is well correlated with the 21\,cm observations, as well as with the
optical reddening $E_{B-V}$ determined by \iras\ \citep{schlegel} as
shown in Figure~\ref{compare}.
The direct comparison of the X-ray and 21\,cm derived columns in
Figure~\ref{nh_nh} shows that the X-ray columns are in general higher
than the 21\,cm ones.  \cite{ab} (AB) showed that below columns of
approximately $0.5 \times 10^{21}$ cm$^{-2}$ the X-ray derived \nh\
column closely matches the 21\,cm value, suggesting that neutral
atomic hydrogen can account for all of the observed column below these
densities.  However, above $0.5 \times 10^{21}$ cm$^{-2}$, the X-ray
column exceeds the 21\,cm column and the contribution from other
hydrogen sources such as molecular hydrogen become important.  This
effect can most easily be seen by plotting the ratio of the X-ray to
21\,cm column as a function of the X-ray column, as we do with our
data in Figure~\ref{nh_nh_ratio}.

\begin{figure}[!t]
\resizebox{\columnwidth}{!}{\rotatebox{+90}{\includegraphics{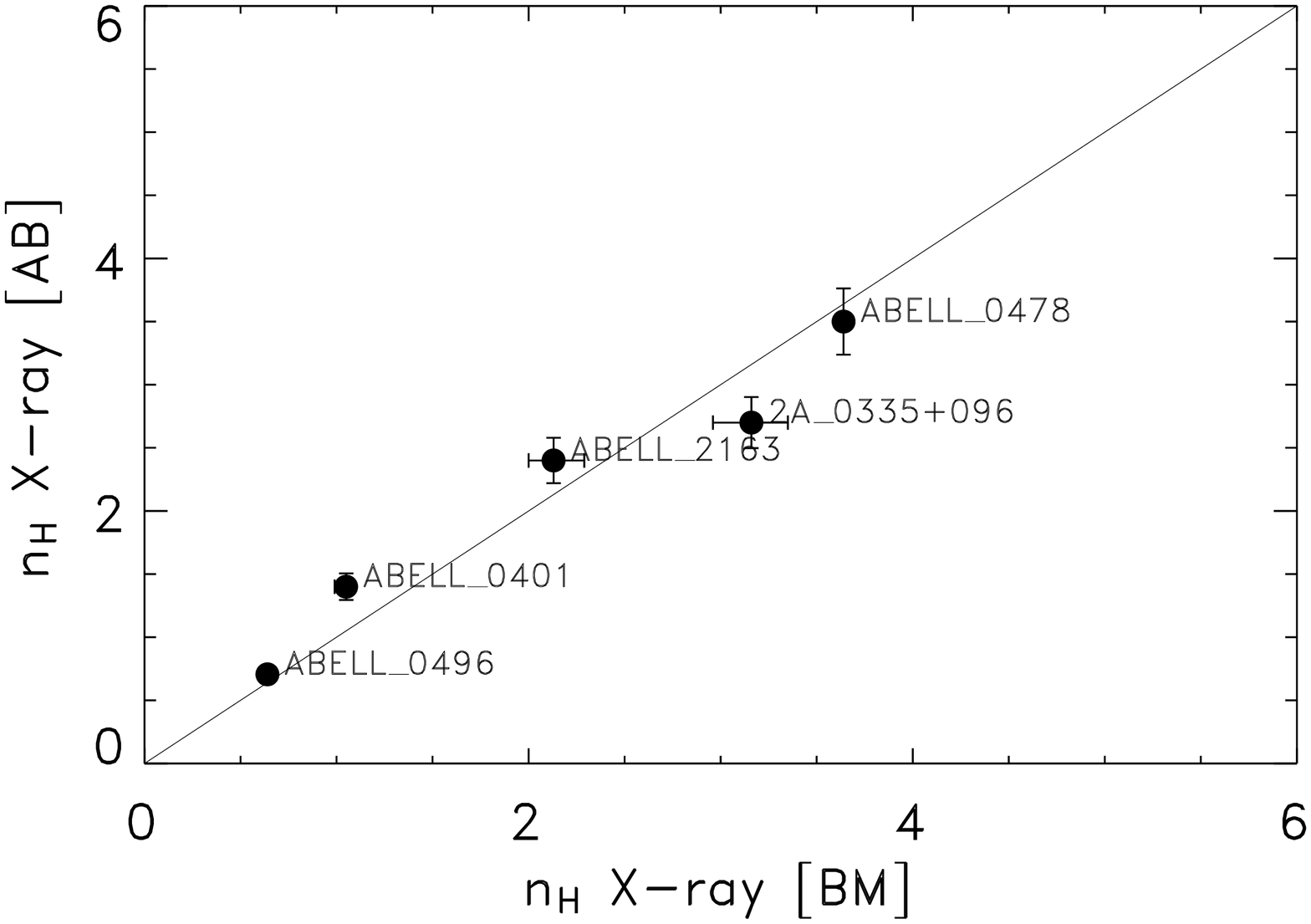}}}
\caption{A comparison of the \cite{ab} X-ray derived \nh\ values to
those in this study [in units of $10^{21}$\,cm$^{-1}$].
\label{ab_wb}}
\end{figure}

While many of our clusters are in the AB list, the better spectral
resolution, broader bandpass, and higher signal to noise data of our
\xmm\ observations are better able to characterize the thermal
emission of the underlying cluster. In Figure~\ref{ab_wb}
we plot our derived X-ray columns against those of AB and find that
they are in good agreement.

Figure~\ref{nh_nh_ratio} indicates that above columns of $0.5 \times
10^{21}$ cm$^{-2}$ the total hydrogen column cannot be accounted
for solely by neutral atomic hydrogen and that other contributors must
be taken into account.  These contributors include ionized hydrogen,
\ion{H}{2}, and molecular hydrogen, \htwo, discussed in the sections
below.

It is also possible that for high columns the 21\,cm derived \nh\
values are incorrect because they assume that the neutral atomic
hydrogen is optically thin.  At higher columns, the gas becomes
sufficiently optically thick to undergo self shielding and
self-absorption.  \cite{strasser} have conducted an
emission-absorption study of \ion{H}{1} in the galactic plane which
measures this effect.  They have found that the 21\,cm columns derived
from emission alone (such as \cite{dl}) require only a $\sim$2\%
correction at columns of $1.0 \times 10^{21}$ cm$^{-2}$ in order
to arrive at the actual galactic hydrogen column.  This correction is
less than the statistical error of our X-ray fits and so we ignore it.

\subsubsection{Ionized Hydrogen}

The contribution of ionized \ion{H}{2} to the total column \nh\ is
small.  Though X-ray photoionization absorption from the ionized
hydrogen is itself not possible, some small amount of absorption from
the associated metals in \ion{H}{2} regions is possible.

AB noted in their study that below columns of $0.5 \times
10^{21}$ cm$^{-2}$ the X-ray total column can be completely
explained solely by contributions from neutral atomic hydrogen as
measured by 21\,cm observations.  \cite{laor} observed the X-ray
spectra of AGN and came to a similar conclusion.  Attempts to
constrain \ion{H}{2} using \iras\ 100\,$\mu$m data by \cite{boulanger}
have also led to low values for \ion{H}{2}, with \cite{kuntz}
suggesting that less than 20\% of the \iras\ emission can be
associated with \ion{H}{2}.

Recently, the \wham\ project has measured \ion{H}{2} emission by
observing at H$\alpha$ \citep{wham}.  They also find that \ion{H}{2}
is not a significant contributor to \nh.

\subsubsection{Molecular Hydrogen and CO Measurements}

CO emission in the radio is known to be correlated with concentrations
of molecular hydrogen.  The correlation is sufficient that a constant
coefficient can be used to estimate the amount of \htwo\ present from
the measured CO emission.  Although the correspondence does not hold
exactly for all densities, (there is significant diffuse \htwo\ at
high galactic latitudes found without substantial CO emission), a
linear correlation \htwo\ = X CO is found at densities high enough for
the \htwo\ to shield the CO from radiation-induced dissociation.  The
value of this so-called X~factor is somewhat controversial and is
thought to vary somewhat depending on the local environment.  However,
\cite{dame} have measured CO emission across the entire galactic plane
and have derived an overall X~factor of $1.8 \times
10^{20}$\,K$^{-1}$km$^{-1}$s\,cm$^{-2}$.

These CO measurements and the X~factor can be used to provide a
measure of the molecular hydrogen content along our lines of sight.
This information can be combined with the X-ray and 21\,cm derived
hydrogen column densities in order to diagnose the hydrogen
composition.  The X-ray derived \nhx\ measures the total column of
hydrogen in all forms, while the 21\,cm column measures the dominant
neutral atomic component.  The CO measurements can then be used to
constrain the proportion of the remaining component that is molecular
hydrogen.

PKS~0745-19 and CIZA~1324 are the only clusters in our sample that
have spatial coordinates that place them within the bounds of the
Dame~\etal\ CO galactic plane survey.  We have extracted CO fluxes of
4.46 and 1.72\,K\,km\,s$^{-1}$ for these lines of sight, which can be
associated with molecular hydrogen columns ($n_{H_{2}}$) of $8.02
\times 10^{20}$\,cm$^{-2}$ and $3.09 \times 10^{20}$\,cm$^{-2}$ using
the X~factor from Dame~\etal.  Equivalent hydrogen columns of \nh\ =
$2.29 \times 10^{21}$\,cm$^{-2}$ and $8.81 \times 10^{20}$\,cm$^{-2}$
are computed using a factor 2.85 for the conversion of \htwo\ to \nh\
as recommended by \cite{wilms}.  

\begin{deluxetable}{lll}
\tablecaption{Hydrogen Column Decomposition\label{table_decomp}}
\tablewidth{0pt}
\tablehead{
\colhead{} & \colhead{PKS~0745-19} & \colhead{CIZA~1324} \\
\colhead{} & \colhead{[$10^{21}$ cm$^{-2}$]} & \colhead{[$10^{21}$ cm$^{-2}$]}
}
\startdata
\nh\ (21\,cm)        & 4.24  & 3.81  \\
$n_{H_{2}}$ (CO)     & 0.802 & 0.309 \\
equivalent \nh\ (CO) & 2.29  & 0.881 \\ \\
\nh\ (21\,cm + CO)\tablenotemark{a}   & $6.53_{5.9}^{7.2}$  & $4.69_{4.2}^{5.2}$  \\
\\ \\
\nh\ (X-ray)\tablenotemark{b}         & $5.6^{5.9}_{5.4}$  & $5.8_{5.2}^{6.5}$  
\enddata
\tablenotetext{a}{The values in sub and superscript are the upper and
  lower range of the error region assuming errors of $\pm 10\%$.}
\tablenotetext{b}{The values in sub and superscript are the extent of the
  90\% confidence region from the X-ray spectral fit.}
\end{deluxetable}

Table~\ref{table_decomp} shows the breakdown of the total hydrogen
column into its components for the two lines of sight towards
PKS~0745-19 and CIZA~1324.  The fourth row shows the sum of the 21\,cm
neutral hydrogen component and the CO derived molecular component, and
the fifth row shows the total column derived from the X-ray
observations.  The errors on the \nh(21\,cm + CO) values come from
assuming a 10\% error.  \cite{ab} show that the error on the 21\,cm
data is approximately 5\%; when combined with the uncertainty in the
CO data and the CO to \htwo\ conversion 10\% is a plausible value.
The error in the \nhx\ value is the 90\% confidence interval from the
X-ray fit.

The results from these two lines of sight show that the \nh(21\,cm +
CO) and \nhx\ values are in fair agreement within the 90\% confidence
regions.

Other evidence supports our claim that the excess absorption in the
X-ray hydrogen column can be ascribed to \htwo.  \cite{federman} show
that at the columns where we begin to see \nhx\ larger than those
measured by 21\,cm measurements, \htwo\ is dense enough to start
shielding itself from incident ionizing radiation that dissociates it
at lower densities.  Also, the \textit{Copernicus} satellite has found
that these are the columns at which \htwo\ starts to become abundant
\citep{savage}.

\subsection{Galactic Oxygen Abundance}

\begin{figure}[!t]
\resizebox{\columnwidth}{!}{\rotatebox{+90}{\includegraphics{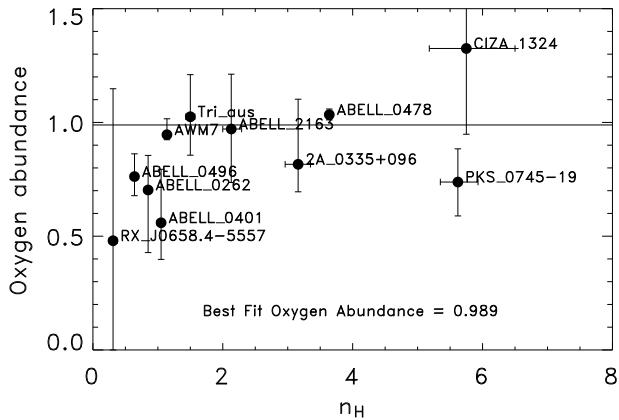}}}
\caption[The galactic oxygen abundance plotted against the X-ray
hydrogen column for 11 lines of sight towards galaxy clusters]{The
galactic oxygen abundance plotted against the X-ray hydrogen column
for 11 lines of sight towards galaxy clusters.  The abundances are
given with respect to the solar value in \cite{wilms}, and the
hydrogen columns are in units of $10^{21}$ cm$^{-2}$.  The best
fit value for the oxygen abundance is O/H $= 0.99 \pm 0.06$ solar.
\label{nh_o}}
\end{figure}

The main result from this work is the X-ray measurement of the
galactic oxygen abundance.  Column~6 of Table~\ref{table_results}
gives our results for the abundance of oxygen using the best fit
galactic absorption model.  These results are plotted against the
X-ray \nh\ from column~5 in Figure~\ref{nh_o}.

The galactic oxygen abundance is uniform for all our lines of sight
and is centered on the solar value.  The best fit value is O/H = 0.99
with a standard deviation of 0.06.  Figure~\ref{nh_o} shows that
there is no trend in the oxygen abundance with increasing hydrogen
column, and that a single value for the oxygen abundance is a
reasonable fit to the individual data points.  These results stand in
contrast to many years of work on metal depletion in the ISM and
support the recent compilation by \cite{jenkins} that shows that
oxygen has little depletion in our galaxy.

\subsection{Oxygen Abundance Variations}

The eleven lines of sight towards the clusters in this study traverse
many different galactic environments.  A diagnostic of the different
regions probed by these sightlines is the ratio of 21~cm \nh\ to \nhx\
that can be used to investigate the density of the observed oxygen.
It is possible that the lines of sight passing through higher density
regions could have different oxygen abundances because of their
proximity to molecular clouds and star forming regions.

In order to try and observe this effect and separate it from the
\textit{a priori} expected abundance variations caused by the galactic
abundance gradient and galactic hydrogen distribution, we use the
Milky Way mass models of \cite{wolfire} and the measurements of the
spatial gradient of the oxygen abundance measured in planetary nebulae
by \cite{henry}.  Our approach is to take the Wolfire~\etal\ density
model for galactic neutral hydrogen, multiply it by the oxygen
abundance in Henry~\etal\ (taking into account the spatial gradient
across the galaxy) and integrate outward from the Sun to determine the
total hydrogen and oxygen columns.  We generate a simple 360 by 180
pixel sky map with pixels spaced every degree in latitude and
longitude. The resulting map indicates the variation of the hydrogen
and oxygen columns across the sky given our simple model, even though
the pixel spacing and map projection emphasize regions at high
latitudes.

The Wolfire~\etal\ model for neutral hydrogen has an exponential
falloff with galactic radius for large radii, and a Gaussian
distribution in the height above the disk.  Also, the center of the
galaxy is low in atomic hydrogen, and Wolfire~\etal\ exclude
hydrogen from the central part of the galaxy:
\[
\Sigma_{\mbox{\ion{H}{1}}}( R ) = \left\{ 
\begin{array}{lcclc}
1.4 R_k - 0.6                         & ( 3 & \le & R_k < &4) \\
5                                     & ( 4 & \le & R_k < &8.5) \\
6.12 \left( R_k/8.5 \right) - 1.12 ~~ & ( 8.5 & \le & R_k < &13) \\
8.24 \,{e}^{-\left(R_k-13\right)/4}   & ( 13 & \le & R_k < &24)
\end{array}
\right. \]
where $\Sigma_{\mbox{\ion{H}{1}}}(R)$ is the \ion{H}{1}
surface density in $M_\odot$\,pc$^{-2}$, $R_k\equiv R/(1$~kpc), and
where we use the conversion 1~$M_\odot$\,pc$^{-2} = 1.25\times
10^{20}$ \ion{H}{1}~cm$^{-2}$.

The Henry~\etal\ spatial oxygen abundance gradient has the
logarithmic abundance going linearly as the galacto-centric radius:
\[
\mbox{O/H}(r) = 10^{(8.97 - 0.037r) - 12.0},
\]
where O/H($r$) is the abundance of oxygen by number with respect to
hydrogen as a function of the galactic radius $r$ in kpc.  The
Henry~\etal\ results are normalized to the solar oxygen abundance
given in \cite{allendeprieto}\footnote{which is nearly identical to
the most recent solar oxygen abundance given in \cite{asplund}} and
measure oxygen abundances in planetary nebulae.  These PN abundances
are consistent with measurements made in \ion{H}{2} regions by
\cite{dehar}.

\begin{figure}[!t]
\resizebox{!}{.25\textheight}{\rotatebox{+90}{\includegraphics{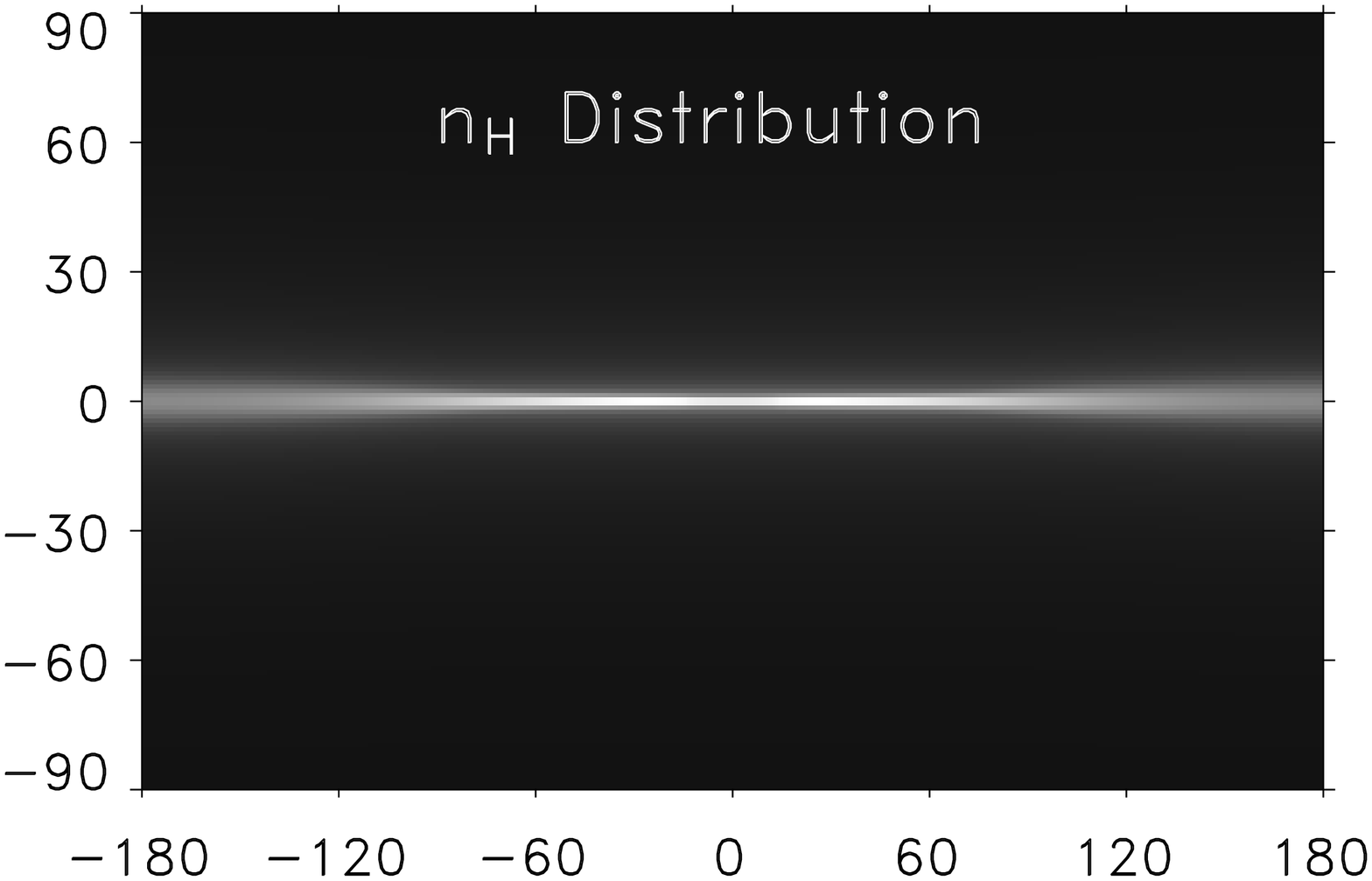}}}
\resizebox{!}{.25\textheight}{\rotatebox{+90}{\includegraphics{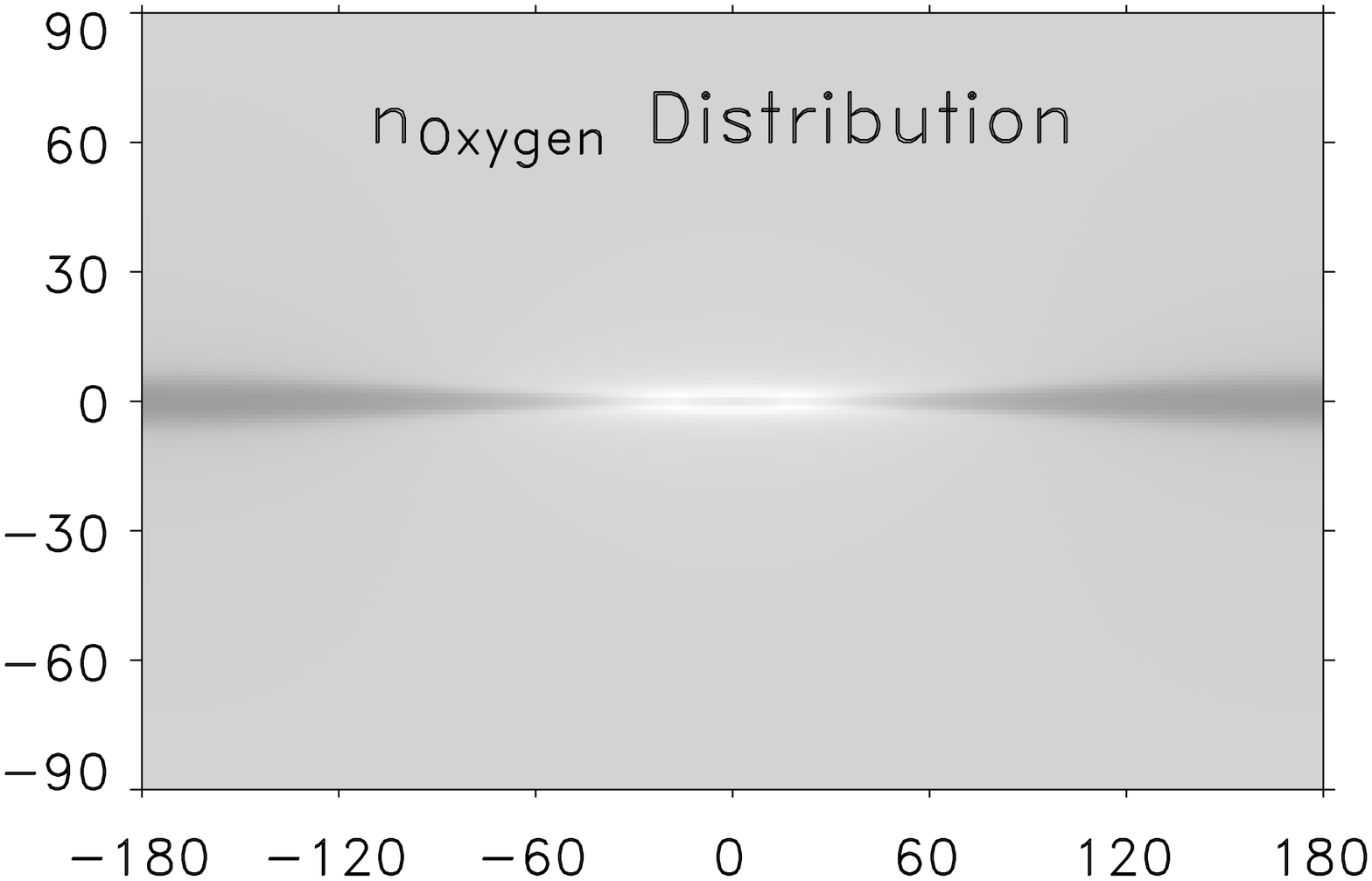}}}
\resizebox{!}{.25\textheight}{\rotatebox{+90}{\includegraphics{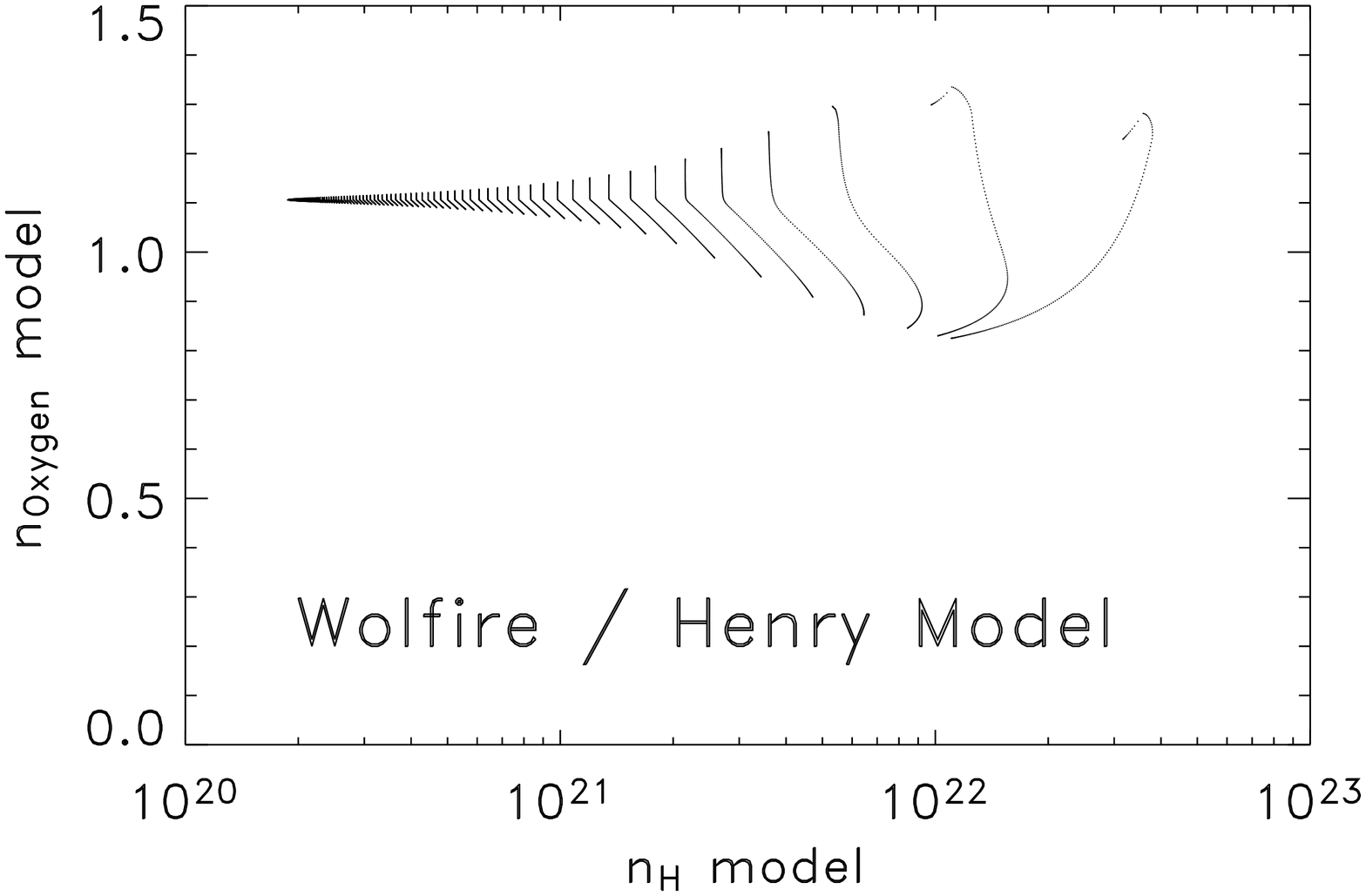}}}
\caption[The distribution of \nh\ and oxygen on the sky as computed by
  integrating along all lines of sight through a model of the
  galaxy]{The distribution of \nh\ and oxygen on the sky as computed
  by integrating along all lines of sight through a model of the
  galaxy.  The first panel shows the integrated neutral hydrogen
  column density integrated from the solar position through the
  galactic density model given by \cite{wolfire}, the second panel
  shows the oxygen abundance derived by incorporating the \cite{henry}
  oxygen gradient into the hydrogen model, and the third panel is a
  scatterplot of the oxygen abundance against the hydrogen column for
  each location in the skymaps.  The lines in the third panel are
  lines of constant galactic latitude, with the low latitudes on the
  right.  The oxygen abundance is very uniform across most of the sky,
  but shows some variation within two degrees of the galactic
  plane.
  \label{model_maps}}
\end{figure}

Figure~\ref{model_maps} shows the hydrogen column across the sky, the
oxygen abundance, and a scatterplot of the oxygen abundance versus the
hydrogen column for sightlines across the entire sky.
This figure shows that we should expect the hydrogen column and oxygen
abundance to be very uniform across most of the sky, except for the
region within a few degrees of the galactic plane.  This is in
agreement with the oxygen results shown in Figure~\ref{nh_o}.

\subsubsection{Spatial Consistency of the Oxygen Abundance}

\begin{figure}[!t]
\resizebox{\columnwidth}{!}{\rotatebox{+90}{\includegraphics{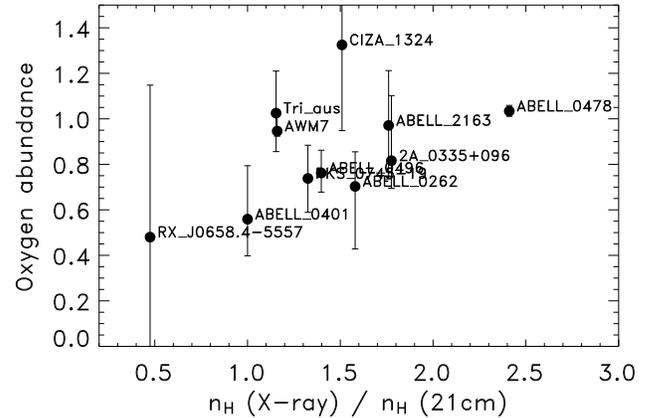}}}
\caption{A plot showing the constancy of the ISM oxygen abundance with
respect to different overdensities of \nhx.  If the high column
density lines of sight can be associated with denser molecular clouds,
this result indicates that the relative abundance of oxygen in these
clouds is not different from its value in other parts of the galaxy.
\label{o_overdense}}
\end{figure}

In order to look for differences in the oxygen abundance related to
galactic environment, we plot the oxygen abundance against the ratio
of the X-ray to 21\,cm \nh\ in Figure~\ref{o_overdense}.
The ratio of the X-ray to 21\,cm derived \nh\ stands in for density:
we expect that any hydrogen seen in the X-ray but not in the radio is
most likely molecular and found in denser clouds.  If the denser
regions have a different oxygen abundance than the more diffuse
regions, we would expect a trend in the plot.  However, when the left
most point with the large error bars is ignored (RX~J0658),
Figure~\ref{o_overdense} indicates that the oxygen abundance shows no
trend with the X-ray to 21\,cm \nh\ ratio, and that the galactic
oxygen abundance is the same in all regions measured.  This disagrees
with the STIS observations reported in \cite{cartledge}, but agrees
with the more recent \fuse\ data of \cite{jensen}.  X-ray observations
towards galactic star forming regions also indicate that these denser
parts of the ISM have abundances similar to those of the local diffuse
ISM \citep{vuong}.

\section{Systematic Errors}

The sytematic errors outweigh the statistical errors in our
measurements.  The main contributors to the systematic error include
the correct determination of the extra edge component in the spectral
fits and the galactic helium abundance.

\subsection{Error in the Extra Edge Determination}

In \S\ref{extraedge} we measured the size of the extra edge necessary
to reconcile the MOS and pn oxygen abundances by looking at high
signal to noise data from the bright sources 3C~273 and the
Coma~cluster and added an extra edge component at oxygen to the MOS
fits so that they would match the pn.  

It is possible that this method does not completely reconcile the MOS
and pn derived abundances for all epochs.  If the magnitude of the
extra edge component is changing with time (e.g., if there is a
progressive buildup of a contaminant in the camera), then we can
expect that the extra edge determined from the calibration sources
will not accurately reflect the value necessary for each of the
cluster observations, which are taken at several different epochs.  We
checked for temporal variations in the extra edge component derived
from 3C273 observations at several epochs but did not find a
significant difference in the data taken at different times.  There
still could be a small time dependent component of the extra edge, but
we argue more generally in the following paragraphs that the
uncertainty in the extra edge component does not have a significant
effect on the measured abundances.

\begin{figure}[!t]
\resizebox{!}{.25\textheight}{\rotatebox{+90}{\includegraphics{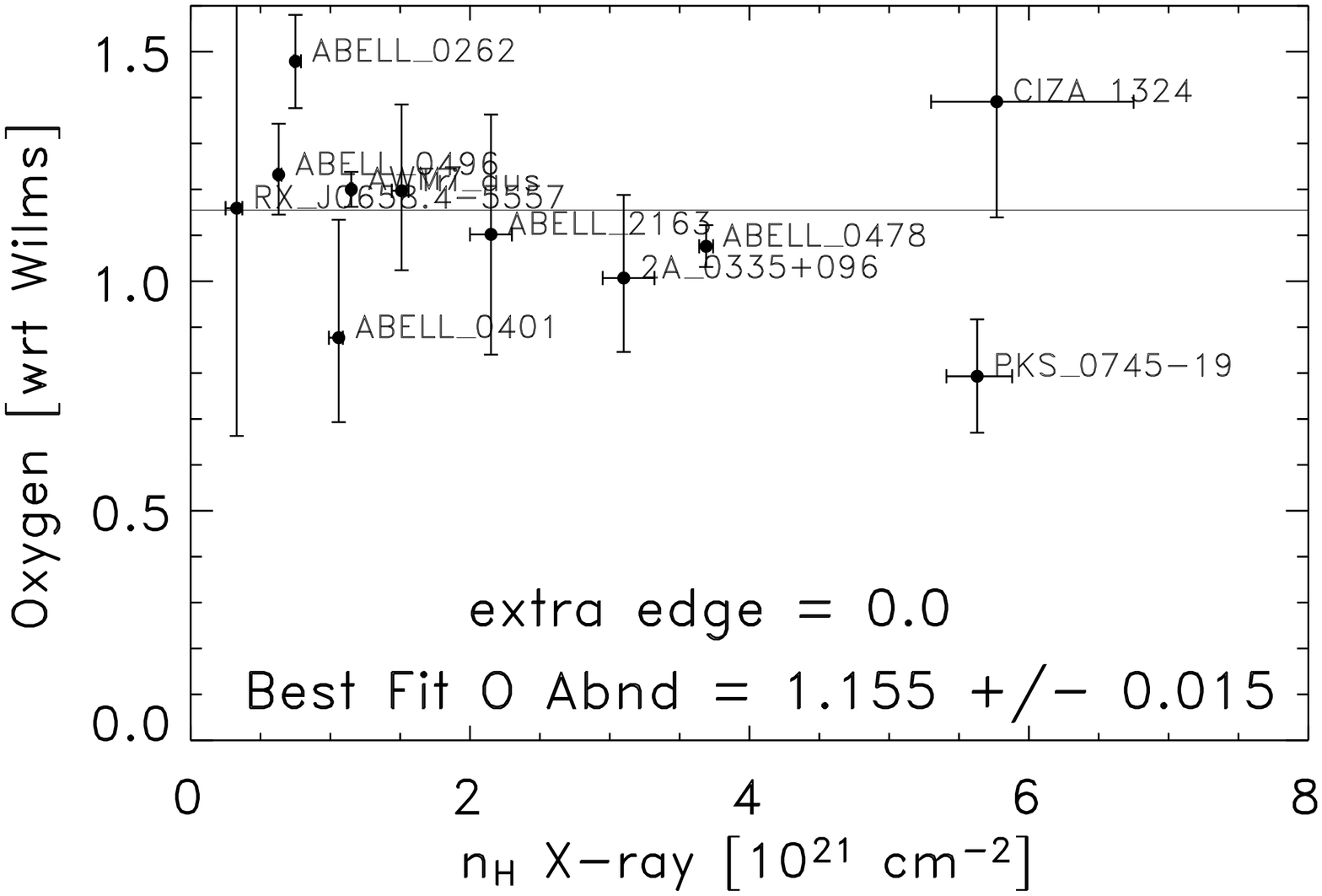}}}
\resizebox{!}{.25\textheight}{\rotatebox{+90}{\includegraphics{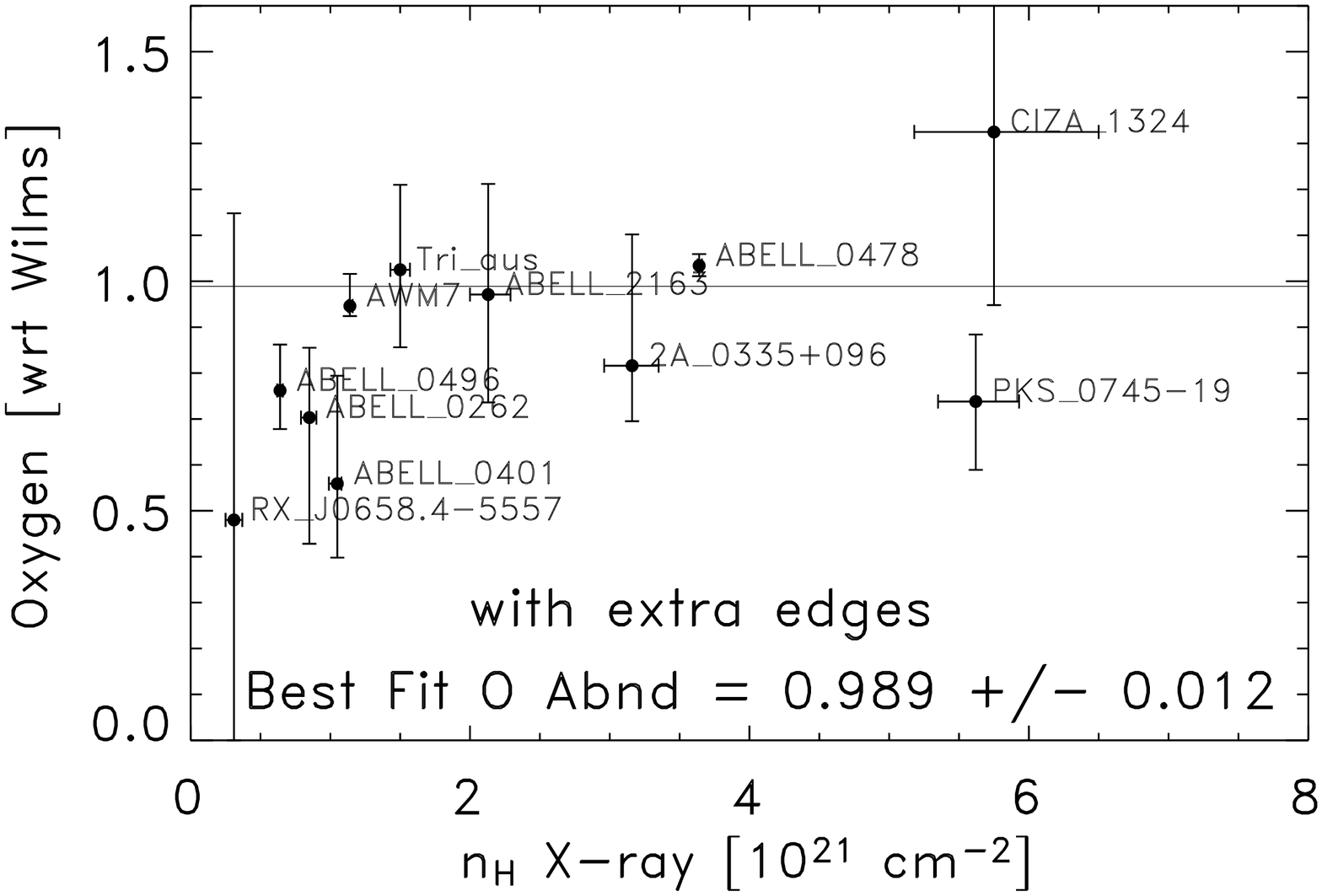}}}
\resizebox{!}{.25\textheight}{\rotatebox{+90}{\includegraphics{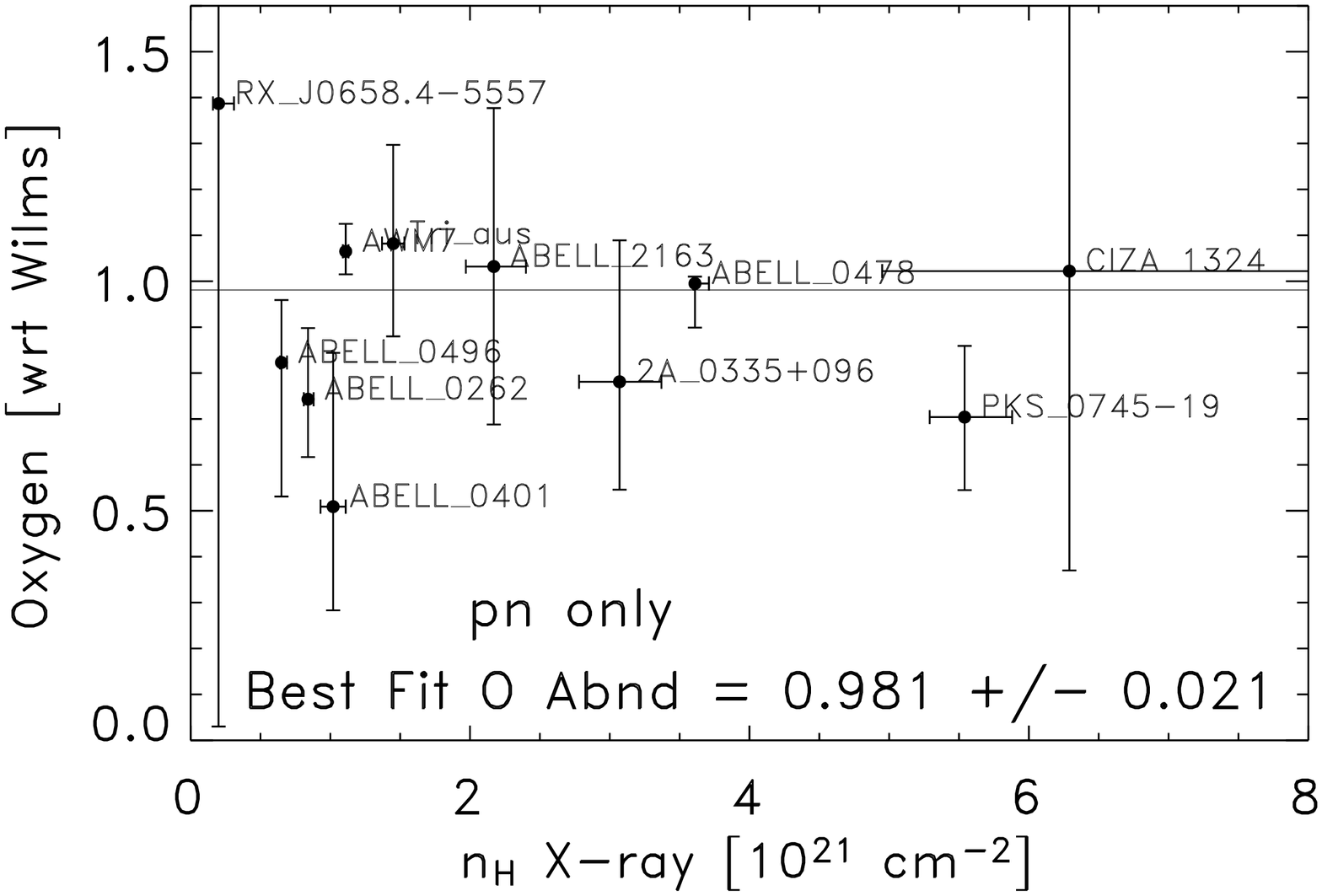}}}
\caption[A comparison of results fit with and without the extra oxygen
  edge model component]{A comparison of results fit with and without
  the extra oxygen edge model component.  The first panel shows the
  results of fitting a model with the extra edge component set to
  zero.  In general, the derived oxygen abundances for the data points
  with the smallest errors are higher than the fits that include the
  extra edge (panel 2) because the fit compensates for the extra
  absorption missing from the instrumental response matrix.  The third
  panel shows the oxygen abundances derived from fits using only data
  from the pn detector in order to avoid the problems with the MOS
  detectors.  The models fit to this data give similar results to the
  models with the extra edge and indicate that the inclusion of the
  extra edge in the model is necessary.
\label{compare_fits}}
\end{figure}

Figure~\ref{compare_fits}
shows two different ways to check this potential problem.  The first
panel of Figure~\ref{compare_fits} shows the ISM oxygen abundances
that result from fitting with the magnitude of the extra edge
component set to zero.  Also, the third panel of
Figure~\ref{compare_fits} shows the oxygen abundances found from
fitting to only the pn data, which does not require the extra edge
component.  

We expect that fits without the extra edge in the model will produce
higher ISM oxygen abundances. Panel one does show that the oxygen
abundances are generally higher without the extra edge than the normal
fits in panel two.

We would also expect the oxygen abundances found from fitting only to
the pn data to be lower than those found with all three detectors and
the extra edge component in the MOS set to zero.  In fact, although
the abundances derived from the pn alone in panel three are slightly
lower than abundances with the extra edge in panel two, the difference
is small.  The main difference is between the data sets fit with and
without the extra edge component (panels one and two).  This
significant difference, and the fact that the models with the edge
match the pn alone models, indicate that the presence of an extra edge
component is necessary for a proper fit and gives good results.

\subsection{Helium Abundance Errors}

Below the oxygen edge at 542\,eV, the absorption due to helium in the
ISM dominates the absorption from all other elements. Above this
energy oxygen dominates the ISM absorption, but the helium component is
still much larger than the hydrogen absorption which is falling off
rapidly approximately as $E^{-3}$.  The galactic helium abundance is
somewhat uncertain, and errors in this value could affect our
abundance measurements because of its strong contribution to the ISM
X-ray absorption at our energies of interest.  If the helium abundance
in our model is too low, the hydrogen component will be increased to
compensate.  This will affect the oxygen abundance, since it is
measured with respect to the hydrogen component.  If the adopted
helium abundance is too high, a similar result occurs, but in the
opposite direction.

We have investigated this effect by fitting our data with two
different helium abundances.  The first data set is fit with the
standard \wilm\ abundances in \xspec, and the second set is fit with
the helium abundance frozen at 80\% of the \wilm\ value.  This value
was chosen because it is representative of the differences between
current helium abundance determinations: The \wilm\ helium abundance
is $9.77 \times 10^{-2}$ by number with respect to hydrogen, and the
helium abundance of \lodd\ is $7.92 \times 10^{-2}$, about a 20\%
difference.

\begin{figure}[!t]
\resizebox{\columnwidth}{!}{\rotatebox{+90}{\includegraphics{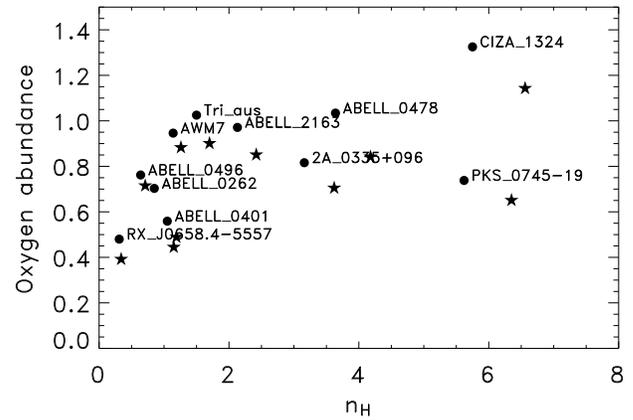}}}
\caption[A demonstration of the effect of an uncertain helium
  abundance on oxygen abundance results]{A demonstration of the effect
  of an uncertain helium abundance on oxygen abundance results.  The
  circular points result from a fit to the data using the solar
  abundances of \cite{wilms}.  The stars are fit using a helium
  abundance of 80\% of the \wilm\ value, as explained in the text.  As
  expected, the lower solar helium abundance drives up the derived
  value of \nh\, which in turn reduces the normalized oxygen
  abundance.  The effect is noticeable, but smaller than the errors on
  the fits (shown in Figure~\ref{nh_o}).
 \label{helium}}
\end{figure}

Figure~\ref{helium}
shows the results from these fits.  The points plotted as circles
(with the cluster name attached) are from the canonical \wilm\ fit,
and the stars are from the low helium fit.  As expected, the lower
helium abundance raises the hydrogen abundance found in the fit and
lowers the oxygen abundance.  However, the effect on the the oxygen
abundance is smaller than the scatter between the multiple cluster
observations.

\section{Summary}

We have used X-ray observations of galaxy clusters to measure the
oxygen abundance and total hydrogen column of the ISM.  Our
measurements of galactic absorption have shown that the X-ray column,
\nhx, is in close agreement with the 21\,cm value for neutral hydrogen
for columns less than approximately $0.5 \times 10^{21}
\mbox{cm}^{-2}$.  Above $0.5 \times 10^{21} \mbox{cm}^{-2}$, the X-ray
column is much higher than the 21\,cm value by up to a factor of 2.5.
This result indicates that there is substantial absorption at high
columns in addition to that provided by neutral hydrogen, and is most
likely from clouds of molecular hydrogen.  Measurements of the
contribution to \nhx\ from H$_2$ derived from CO observations show
that the molecular hydrogen column density makes up for the observed
difference.

We also measure the ISM oxygen abundance by observing the K-shell
photoionization edge at 542\,eV\@.  After taking into account
calibration problems in the EPIC MOS detectors on \xmm, we find that
the galactic oxygen abundance is consistent with the most recent solar
values.  Previously, oxygen abundance measurements have suggested that
oxygen is depleted in the ISM because gas phase measurements have
shown the abundance to be lower than the adopted solar value.
However, the accepted solar value has recently decreased as a result
of better modeling of solar spectra \citep{asplund}.  Our measurement
in conjunction with the recent solar value shows that oxygen is not
depleted in the ISM, in agreement with \cite{jenkins}.  We also find
that the oxygen abundance is uniform across an order of magnitude in
\nhx, suggesting similar compositions for higher density regions in
the ISM and more diffuse regions, in agreement with the conclusions of
\cite{vuong}.

\acknowledgements

The authors would like to thank H.\@ Ebeling and T.\@ Furusho for
sharing their \xmm\ data for CIZA~1324 and AWM~7 before the public
release dates, and K.\@ Kuntz for many valuable discussions.  This
work has made use of data from the High Energy Astrophysics Science
Archive Research Center (HEASARC), provided by NASA's Goddard Space
Flight Center as well as data from the NASA/IPAC Extragalactic
Database (NED) which is operated by the Jet Propulsion Laboratory,
California Institute of Technology, under contract with the National
Aeronautics and Space Administration.


\end{document}